\documentclass[fleqn,twoside]{article}
\usepackage{espcrc2}

\usepackage{graphicx}
\usepackage[figuresright]{rotating}

\newcommand{\AmS}{{\protect\the\textfont2
  A\kern-.1667em\lower.5ex\hbox{M}\kern-.125emS}}

\title{ Detecting Solar  Neutrino Flare in Megaton and $km^3$ detectors}

\author{Daniele Fargion, Paola Di Giacomo\address[un1-INFN]{Physics Department, Rome Univ.1 , Sapienza \\
        Ple A. Moro 2, 00185, and INFN, Rome,Italy}%;
       }

\begin{document}

\begin{abstract}
To foresee a solar flare neutrino signal we infer its upper and lower bound.
The upper bound was derived since a few years by general energy equipartition arguments on observed solar particle flare.
The lower bound, the most compelling one for any guarantee neutrino signal,
is derived by most recent records of hard Gamma bump due to solar flare on January $2005$ (by neutral pion decay).
Because neutral and charged  pions (made by hadron scattering in the flare) are born on the same foot, their link is compelling: the observed gamma flux \cite{ref2} reflects into a corresponding one for the neutrinos, almost one to one. Moreover while gamma photons might be absorbed (in deep corona) or at least reduced inside the  flaring plasma, the secondaries neutrino are not. So pion neutrinos should be even more abundant than gamma ones. Tens-hundred MeV neutrinos may cross undisturbed the whole Sun, doubling at least their rate respect a unique solar-side for gamma flare. Therefore we obtain minimal bounds opening a windows for neutrino astronomy, already at the edge  of present but \emph{quite within near future  Megaton neutrino detectors}. Such detectors are considered mostly to reveal cosmic supernova background or rare Local Group  (few Mpc) Supernovas events \cite{ref3}. However  rarest (once a decade), brief (a few minutes) powerful solar neutrino "flare"  may shine and they may overcome by two to three order of magnitude the corresponding steady atmospheric neutrino noise on the Earth, leading in largest Neutrino detector at least to one or to  meaning-full few events clustered signals. The voice of such a solar anti-neutrino flare  component at a few tens MeVs may induce an inverse beta decay over a
$vanishing$  anti-neutrino solar background. Megaton or even   inner ten Megaton Ice Cube detector at ten GeV threshold
may also reveal  traces in hardest energy of solar flares. Icecube , marginally, too. Solar neutrino flavors may shine light on neutrino mixing
angles. Not only on orbit satellites but even human  astronauts in Space   may exploit underground neutrino detectors for the prompt alert on (otherwise) fast and maybe lethal solar explosions.
\vspace{1pc}
\end{abstract}
% typeset front matter (including abstract)
\maketitle
%%%%%%%%%%%%%%%%%%%%%%%%%%%%%%%%%%%%%%%%%%%%%%%%%%%%%%%%%%%%%%%%%%%%%%%%%%%%%%%%
\section{ Solar Neutrino Flare Astronomy at edge}
During largest solar flare, of a few minutes duration,  the
particle flux escaping the corona eruption and hitting later on
the Earth, is 3-4 order of magnitude above the common atmospheric
CR background . If the flare particle interactions on the Sun corona is taking place as efficiently as  in terrestrial atmosphere, than their secondaries
by charged pions and muons decays, are leading to a neutrinos
fluency on Earth comparable to one day terrestrial atmospheric
neutrino activity (upper Bound). One therefore may expect a prompt increase of
neutrino signals of the order of one day integral events made by
atmospheric neutrinos \cite{ref0}. In present neutrino detectors the signal is
just on the edge, but as long as the authors know, it has been
never revealed \cite{ref1}.  Sun density at the flare corona might be diluted
and pion production maybe consequently suppressed (by a factor
(0.1-0.05)). This may be the reason for the SK null detection. Indeed the low Gamma signals recently reported \cite{ref2} confirm this suppressed signal, but just at the detection edge (See Fig.\ref{01},\ref{02}). Unfortunately the neutrino signal at  hundred MeV energies is rare while the one at ten MeV or below is polluted by Solar Hep neutrinos.  The expected signal is dominated by $10-30$ MeV neutrinos, that  might be greatly improved  by   anti-neutrino component via Gadolinium presence in next SK detectors \cite{ref1}.  Our earliest (2006) upper limit
estimate for  October - November $2003$ solar flares \cite{ref0} and  the recent January $20th$ 2005 \cite{ref1} exceptional flare were  leading to signal
near unity for  Super-Kamiokande and to a few events above   unity for Megaton detectors (See Fig.\ref{01},\ref{02}). Our present estimate based on a lower (gamma) bound of neutrino signal, while below the upper ones, still confirms the order of magnitude and the near edge discovery (See Fig.\ref{01},\ref{02}). The recent peculiar solar flares as the October-November $2003$ and January $2005$ \cite{ref2} were source of high energetic charged particles: large fraction of these {\itshape{primary}} particles, became a source of both neutrons \cite{ref1} and {\itshape{secondary}} kaons, $K^{\pm}$, pions, $\pi^{\pm}$ by their particle-particle spallation on the Sun surface \cite{ref0}. Consequently,
$\mu^{\pm}$, final secondaries muonic and electronic neutrinos and anti-neutrinos,
${\nu}_{\mu}$, $\bar{\nu}_{\mu}$, ${\nu}_{e}$, $\bar{\nu}_{e}$,
$\gamma$ rays, are released by the chain reactions $\pi^{\pm}
\rightarrow \mu^{\pm}+\nu_{\mu}(\bar{\nu}_{\mu})$, $\pi^{0}
\rightarrow 2\gamma$, $\mu^{\pm} \rightarrow
e^{\pm}+\nu_{e}(\bar{\nu}_{e})+ \nu_{\mu}(\bar{\nu}_{\mu})$
occurring on the sun atmosphere. There are two different sites for
these decays (see \cite{ref0}): A brief and sharp
solar flare, originated within the $solar$ corona itself  and a diluted
and delayed $ terrestrial$ neutrino flux, produced by late flare
particles hitting the  Earth's  atmosphere. This latter delayed
signal is of poor physical interest, like an inverse missing signal during the E. Forbush  phase. The main and first {\itshape{solar}} flare neutrinos reach the Earth with a well defined directionality and within a narrow time range. The
corresponding average energies $<E_{{\nu}_{e}}>$,
$<E_{{\nu}_{\mu}}>$  ( since  low solar corona densities)
suffer negligible energy loss: $<E_{{\nu}_{e}}> $ $\simeq$ $50
MeV$, $<E_{{\nu}_{\mu}}> \simeq$ 100 $\div$ 200 MeV. The opposite
occur to downward flare. In the simplest approach, the main source
of pion production is $p+p\rightarrow {{\Delta}^{++}}n\rightarrow
p{{\pi}^{+}}n$; $p+p\rightarrow{{{\Delta}^{+}}p}^{\nearrow^{p+p+{{\pi}^{0}}}}_{\searrow_{p+n+{\pi}^{+}}}$
at the center of mass of the resonance ${\Delta}$ (whose mass
value is ${m}_{\Delta}=1232$ MeV). As  a first approximation and
as a useful simplification after the needed boost of the
 secondaries energies  one may assume that the total pion $\pi^+ $ energy
is equally distributed, in average, in all its final remnants:
($\bar{\nu}_{\mu}$, ${e}^{+}$, ${\nu}_{e}$,
${\nu}_{\mu}$):${E}_{{\nu}_{\mu}} \geq {E}_{{\bar{\nu}_{\mu}}}
\simeq {E}_{{\nu}_{e}} \simeq \frac{1}{4}{E}_{{\pi}^{+}}$. Similar
nuclear reactions (at lower probability) may also occur by
proton-alfa scattering leading to: $p+n\rightarrow
{{\Delta}^{+}}n\rightarrow n{{\pi}^{+}}n$; $p+n\rightarrow
{{{\Delta}^{o}}p}^{\nearrow^{p+p+{{\pi}^{-}}}}_{\searrow_{p+n+{\pi}^{o}}}$.
Here we neglect the ${\pi}^{-}$ additional role due to the flavor
mixing and the dominance of previous reactions ${\pi}^{+}$. To a
first approximation the flavor oscillation will lead to a
decrease in the muon component and it will make the electron
neutrino component a bit harder. Indeed the oscillation length (at
the energy considered) is small with respect to the Earth-Sun distance:
 $ L_{\nu_{\mu}-\nu_{\tau}}=2.48 \cdot10^{9} \,cm \left(
 \frac{E_{\nu}}{10^{9}\,eV} \right) \left( \frac{\Delta m_{ij}^2
 }{(10^{-2} \,eV)^2} \right)^{-1} \ll D_{\oplus\odot}=1.5\cdot
 10^{13}cm$.  While at the birth place the neutrino
fluxes by positive charged pions $\pi^+$ are
$\Phi_{\nu_e}$:$\Phi_{\nu_{\mu}}$:$\Phi_{\nu_{\tau}}$ $= 1:1:0$,
after the mixing assuming a democratic number redistribution we
expect $\Phi_{\nu_e}$:$\Phi_{\nu_{\mu}}$:$\Phi_{\nu_{\tau}}$ $=
(\frac{2}{3}):(\frac{2}{3}):(\frac{2}{3})$. Naturally in a more
detailed balance the role of the most subtle and hidden parameter
 (the neutrino mixing  $\Theta_{13}$)  may be deforming the present
averaged  flavor balance. On the other side for the anti-neutrino
fluxes we expect at the birth place:
$\Phi_{\bar\nu_e}$:$\Phi_{\bar\nu_{\mu}}$:$\Phi_{\bar\nu_{\tau}}$
$= 0:1:0$ while at their arrival (within a similar democratic
redistribution): $\Phi_{\bar\nu_e}$:$\Phi_{\bar\nu_{\mu}}$:$\Phi_{\bar\nu_{\tau}}$
$= (\frac{1}{3}):(\frac{1}{3}):(\frac{1}{3})$.
This neutrino  flux, derived by gamma one, hold $100$ s  duration and it is larger by
two order of magnitude over the atmospheric one.
\section{Detections in SK,  Megaton, ICECUBE}
Therefore in present paper we conclude  that, even if SK just marginally missed the Solar neutrino flares (See Fig.\ref{01},\ref{02}),  Hyper Kamiokande, HK, or Megaton, Titand \cite{ref3}, detectors should discover neutrino signals quite above threshold edge (See Fig.\ref{03}). Inner Ice-Cube $ 0.01 km^3$ may reach the threshold if ten GeV neutrinos maybe observed. Finally full $km^3$ ICECUBE, being detecting at highest $50$ GeV may or may-not reveal tens GeV neutrinos (as observed gamma found in Milagro ones) just at edge (See Fig.\ref{04}).Solar Neutrino $\nu_e$, $\nu_{\mu}$ Flare After their
mixing in two different upper bounds considered in early papers \cite{ref0},\cite{ref1},derived by Solar flare energy equipartition  in two solar corona density target. Antineutrino exhibit  similar fluxes over   a noise-free solar antineutrino background \cite{ref1}. Here the signal is derived, by  pions connection and gamma detection of Solar flare on 20 January $2005$,\cite{ref2} in SK. While upper bounds appears  well within SK detection,  the observed lower gamma bound quite below, is just (marginally) out of the SK detection thresholds. This is consistent and it  explains the solar flare neutrino absence (SK private communication). The solar  flare duration and power (here and in next figures), from where we derived  the expected neutrino signals are assumed about $100$ s long  as powerful as the observed $20$ Jan. $2005$ event. The vertical arrows among the arcs describes our estimated solar flare neutrino flux windows. The vertical dotted blue lines is related to SK cut-threshold via neutrino-nucleon  CC, whose cross-section is quite above the electron mass. The solar
 neutrino noise rules lowest energies, but not the tens MeV energy band.
 In analogy the green dashed line , along $\nu_{\mu}$ , shows the muon neutrino thresholds,(near GeV), just out SK detection. The different dotted lines split on from  electrons, taus and the most inclined green ones
for muons: indeed their penetrability is  growing above the SK detector size, tens meters size, increasing their detections. Red dot-dashed line shows rarest  $\nu_{\tau}$ appearance by  mixing  and by large Sun-Earth distances. The same phenomenon is difficult for atmospheric neutrino because of shorter Earth size.
%\clearpage

\end{document}